\documentclass{article}

\usepackage{amssymb,amsmath}
\usepackage{fullpage}
\usepackage{graphicx}
\usepackage{amsmath}		
\usepackage[margin=1.0in]{geometry}
\usepackage{setspace}
\usepackage{color}
\usepackage{fancyhdr}
\usepackage{collcell}
\usepackage{datatool}
\usepackage{environ}
\usepackage{latexsym}
\usepackage{amssymb}
\usepackage{epsfig,amsmath,graphics}
\usepackage{epstopdf}
\usepackage{verbatim}
\usepackage{wasysym}
\usepackage{feynmp-auto}
\usepackage{authblk}

\usepackage[utf8]{inputenc}

\title{Storage Ring Probes of Dark Matter and Dark Energy}
\date{\today}

\author[1]{Peter W.~Graham}	
\author[2]{Selcuk Hac\i\"omero\u glu}
\author[3]{David E.~Kaplan}
\author[4]{Zhanibek Omarov}
\author[3]{Surjeet Rajendran}
\author[2,4]{Yannis K.~Semertzidis}
\affil[1]{Stanford Institute for Theoretical Physics, Department of Physics, Stanford University, Stanford, CA 94305, USA}
\affil[2]{Center for Axion and Precision Physics Research, Institute for Basic Science, Daejeon 34051, Republic of Korea}
\affil[3]{Department of Physics \& Astronomy, The Johns Hopkins University, Baltimore, MD  21218, USA}
\affil[4]{Department of Physics, Korea Advanced Institute of Science and Technology, Daejeon 34141, Republic of Korea}

\usepackage{enumitem}
\begin{document}

\maketitle

\begin{abstract}
We show that proton storage ring experiments designed to search for proton electric dipole moments can also be used to look for the nearly dc spin precession induced by dark energy and ultra-light dark matter. These experiments are sensitive to both axion-like and vector fields. Current technology permits probes of these phenomena up to three orders of magnitude beyond astrophysical limits. The relativistic boost of the protons in these rings allows this scheme to have sensitivities comparable to atomic co-magnetometer experiments that can also probe similar phenomena. These complementary approaches can be used to extract the micro-physics of a signal, allowing us to distinguish between pseudo-scalar, magnetic and electric dipole moment interactions. 

\end{abstract}

\tableofcontents

\section{Introduction}
A variety of cosmological and astrophysical measurements have established that nearly 95\% of the energy density in the universe today is in dark energy and dark matter. The nature of these cosmic fluids remain a mystery. Since these fluids form a preferred (cosmic) background, a generic way to search for the properties of these fluids is to look for the effects of the motion of standard model particles against this cosmic background `ether'. Experiments that test Lorentz invariance can thus be reinterpreted as searches for interactions between the standard model and these dark fluids \cite{Pospelov:2004fj}. 

 Spin precession experiments \cite{Romalis:2013kkt, Bennett:2008cpt, Brown:2010dt, Graham:2017ivz,Gemmel:2010ft} are a canonical test of Lorentz invariance. These look for an anomalous precession of a spin caused by an interaction between a spin moving against a background classical field. These experiments are a particularly well motivated way to search for ultra-light dark matter candidates such as axions and axion-like-particles, which naturally possess such spin-precession inducing interactions \cite{Graham:2013gfa}. Moreover, unlike dark matter where ultra-light particles are simply a possible class of dark matter candidates, any viable dark energy candidate that is not a cosmological constant has to be a ultra-light classical field. Axion-like interactions that induce spin precession are also a natural expectation of such dark energy candidates, particularly when those candidates can actually solve the cosmological constant problem \cite{Graham:2019bfu, Graham:2017hfr}. 

Spin precession experiments that search for axion dark matter \cite{Budker:2013hfa} are largely focused on axions with a mass larger than $\sim$ 1 Hz. In this regime, many technical sources of noise (such as vibrations and magnetic shielding) are more easily ameliorated. There is however considerable scientific motivation to develop techniques that can search for such signals at much lower frequencies. This is important since the observational limit on the mass of dark matter is as low as $10^{-7}$ Hz, wherein dark matter in this mass range gives rise to an essentially DC signal in an experiment. Moreover, the spin precession induced by dark energy will also be a DC signal since any change to the frequency of this signal has to be comparable to the Hubble scale $\sim 10^{-18}$ Hz.   The challenges of a DC spin precession signal are currently combated through the use of atomic co-magnetometers \cite{Romalis:2013kkt, Brown:2010dt, Graham:2017ivz,Gemmel:2010ft}, where the relative precession between two species in a medium is used to cancel out many sources of technical noise. 

In this paper, we propose the use of storage rings as an alternate technique to search for such spin precession. Storage rings are devices where a relativistic beam of particles (for concreteness, we will consider protons in this paper) are stored for significant periods of time by means of electric and/or magnetic fields. These rings possess a magic momentum wherein the precession of a spin in the electromagnetic fields of the ring is compensated by the motion of the particle around the ring thus maintaining a fixed orientation of the spin relative to the direction of motion of the particle. By measuring the spin of the particles, these storage rings can search for anomalous precession. These rings have been used to search for fundamental electric dipole moments (EDMs) \cite{Farley:2003wt, Anastassopoulos:2015ura}. The notion of the magic momentum was first invented at the third CERN muon $g-2$ experiment, where the muon spin precession frequency is not affected by the presence of the electric focusing system, while its $g-2$ frequency can be measured with high accuracy in a highly uniform B-field.\cite{Bailey:1979np} 

Inspired by these designs, we evaluate the feasibility of storage rings as a way to search for ultra-light dark matter and dark energy. We will show that storage rings can have sensitivities comparable to atomic co-magnetometer techniques for pseudo-scalar interactions. For vector backgrounds, due to the relativistic nature of the beam, these rings have enhanced sensitivity to magnetic dipole interactions and can thus distinguish between electric and magnetic dipole interactions. Storage ring techniques are thus complementary to atomic co-magnetometer searches --- the combination of both techniques can be used to extract the underlying nature of any new physics discovered in such experiments. The rest of this paper is organized as follows: in section \ref{sec:theory}, we present a theoretical overview of the signals of dark energy and dark matter in a storage ring. In section \ref{sec:storagering}, we discuss the storage ring setup and evaluate the sensitivity of this approach. Following this, we discuss experimental backgrounds and ways to ameliorate them to the required levels in section \ref{sec:Experimental setup and backgrounds overview}. We conclude in section \ref{sec:discussion}, where we also compare and contrast this approach to atomic co-magnetometer searches in greater detail.

\section{The Theory of the Effect}
\label{sec:theory}

For the purposes of this article, we are interested in extremely light (pseudo-)scalar fields coupled to protons.  Such a coupling can come from an underlying coupling to quarks, {\it e.g.},
\begin{equation}
    {\cal L}\supset g_{aqq} {\partial_\mu a} \bar{q}\gamma^\mu \gamma_5  q \rightarrow g_{aNN}{\partial_\mu a} \bar{p}\gamma^\mu \gamma_5  p
\end{equation}
Where $a$ is the light spinless field, and $q$ is a light quark and $p$ is the proton. Such quark couplings can come from spontaneous breaking of a global symmetry at high scales.  Assuming such a symmetry is anomaly-free with respect to QCD, the 
'axion', $a$, does not receive an instanton-induced mass and will remain naturally light.

In the rest frame of the proton, this term generates the effective Hamiltonian term:
\begin{equation}
H \ni - g_\text{aNN} \vec{\nabla} a \cdot \vec{\sigma}
\end{equation}
where $\sigma$ is the proton spin.  Thus, the gradient of the axion field couples to the spin of the proton like a magnetic field.  The physical effect is the precession of the proton spin around the vector $\vec{\nabla} a$.  The size of the gradient of the axion field is set by the axion momentum: $\vec{\nabla} a \sim \gamma m_a \beta a$ where $m_a$ is the axion mass and $\beta$ is its velocity in the proton rest frame and $\gamma$ is the relativistic gamma factor.  

Here is where the storage ring is interesting.  A laboratory experiment is moving at a velocity $\beta \sim 10^{-3}$ relative to the virial velocity of dark matter or the rest frame of dark energy, either of which $a$ could be a component.  On the other hand, in a storage ring experiment at relativistic velocities, we get an automatic enhancement with $\beta\sim{\cal O}(1)$.  The $\gamma$ factor however cancels due to time dilation, as the integration time is measured in the lab frame.

\subsection{Dark Energy}
\label{sec:DE}

In the case of dark energy, we imagine a field $a$ rolling down a shallow potential with a non-zero $\partial_t a$, homogeneous in the rest frame of the CMB.  Taking the kinetic energy of $a$ to be a fraction $\epsilon$ of the dark-energy energy density $\rho_\text{DE}$, we have (in the proton rest frame):
\begin{equation}
    \partial_\mu a = \{ \gamma  \sqrt{2\epsilon\rho_\text{DE}}, \gamma \vec{\beta}  \sqrt{2\epsilon\rho_\text{DE}} \}
\end{equation}
and thus the proton's spin will to couple the axion gradient analogous to how a  magnetic field couples to its magnetic moment.  A proton spin not aligned with the direction of the velocity will precess around $\vec{\beta}$ at a precession frequency of $\nu_{prec} \simeq g_{\rm aNN} \beta  \sqrt{2\epsilon\rho_\text{DE}}/2\pi$ in cycles per second in the lab frame.  Note that for a proton traveling in a straight line, the effect always rotates the spin around the same direction around the velocity, and thus the effect adds over long integration times.  Thus, after a time $T$, the precession angle would be $\Delta\theta_{prec} \simeq g_{\rm aNN} \beta  \sqrt{2\epsilon\rho_\text{DE}}T$.  The size of the effect will be set by the coupling $g_\text{aNN}$, the experimental value of dark energy density is $\rho_\text{DE}  \sim (2\times 10^{-3} {\rm \; eV})^4$, and the fraction of dark energy in the kinetic rolling $\epsilon$.  The current bound on $\epsilon$ can be loosely extracted from the Planck collaboration's fit  to dynamical dark energy: $\epsilon\simeq (w+1)/2 < 0.06 (0.20)$ at $1\sigma (2\sigma)$, where $w$ is the dark energy equation of state measured today \cite{Aghanim:2018eyx}.

\subsection{Dark Matter}
\label{sec:DM}
\subsubsection{Pseudoscalars}
In the case of $a$ dark matter, the main difference with dark energy is the fact that the $a$ field value (and thus the time derivative) oscillates in time as 
$a \sim \frac{\sqrt{2 \rho_\text{DM}}}{m_a} \sin(m_a t)$, 
where $\rho_\text{DM}$ is the local dark matter energy density.  
From time $t$ to $dt$, the proton spin precesses an amount 
$d\theta_{prec} \simeq g_{\rm aNN} \beta \sqrt{2 \rho_\text{DM}} \cos(m_a t) dt$.
Thus, in the case of dark matter, the precession angle after a time $T$ would be
\begin{equation}
\Delta\theta_{prec} \simeq \frac{g_{\rm aNN} \beta \sqrt{2 \rho_\text{DM}}}{m_a} \sin(m_a T))
\end{equation}
in the lab frame.  For $m_a\ll 1/T$, the effect again adds coherently, whereas for $m_a\gtrsim 1/T$, the precession angle oscillates in time.
\\
\subsubsection{Vectors}
We also note that spin precession of the proton would also result if the  dark matter were a light vector, $A_\mu '$.   Consider the magnetic dipole moment and electric dipole moment operators, which couple the vector (hidden photon) field to protons: 
\begin{equation}
\mathcal{L} \supset g_{\rm MDM}' F^{\prime}_{\mu\nu}\bar{p}\,\sigma^{\mu\nu}p +  g_{\rm EDM}' F^{\prime}_{\mu\nu}\bar{p}\,\gamma^5 \sigma^{\mu\nu}p 
\end{equation}
where $F^{\prime}_{\mu\nu} = \partial_{\mu}A^{\prime}_{\nu} - \partial_{\nu}A^{\prime}_{\mu}$. Under the same assumption as in the axion case that the hidden (dark) photon field can be modeled as a classical wave, we can estimate the amplitude of the oscillating dark electric field as $E_{\rm lab}^{\prime}\simeq \sqrt{\rho_{\rm{DM}}}$ in the rest frame of the CMB, while dark magnetic field is velocity-suppressed in this frame.  In the frame of a proton traveling at a relativistic velocity $\vec{\beta}$, the dark magnetic field is $\vec{B}'= \vec{\beta}\times \vec{E}'$, relative to the dark electric field in the same frame.  In this frame, the spin of the proton will precess around the B- and E-fields due to the MDM and EDM operators respectively.

For a proton traveling in a circular ring, a dark E-field in the lab frame, perpendicular to the plane of the ring generates a dark B-field  pointing radially.  For the MDM operator, a spin locked in the direction of the velocity $\vec{\beta}$ will precess out of the ring's plane.  An E-field component in the ring's plane will  cause precession in the plane of the ring that oscillates.  Such precession will thus cancel over each round trip of the proton.  Thus, if the dark electric field is pointing at an angle $\phi$ with respect to the ring plane's normal, a proton spin initially parallel to velocity will have a precession angle after a time $T$ of
\begin{equation}
\Delta\theta_{prec} \simeq \frac{g_{\rm MDM}' \beta \sqrt{\rho_\text{DM}}}{m_{A'}} \cos{\phi}\sin(m_{A'} T),
\end{equation}
where we have ignored the in-plane oscillating precession by assuming an integer number of trips around the ring.  The $\cos{\phi}$ picks out the normal component of the E-field, and $m_{A'}$ is the dark photon mass.

Aligning the spin parallel with the velocity, the EDM operator will cause a similar precession due to a perpendicular dark E-field, but this time in the plane of the ring.  One can get precession out of the plane due to a co-planar dark E-field.  If the spin is {\it kept in a fixed planar direction in the lab frame}, then the dark planar E-field will produce a precession out of the plane after a time $T$
\begin{equation}
\Delta\theta_{prec} \simeq \frac{g_{\rm EDM}' \sqrt{\rho_\text{DM}}}{m_{A'}} \sin{\phi}\sin(m_{A'} T),
\end{equation}
where $\phi$ again is the angle between the dark E-field and the plane normal.

An interesting aspect of this signal is that its magnitude should have a daily modulation as the plane of the ring evolves with the Earth's and the angle $\phi$ oscillates with a 24 hour period.  This will be a second modulation on top of the dark matter's at angular frequency $m_{A'}$

\section{A Storage Ring Experiment}
\label{sec:storagering}

The spin precession caused by the axion wind effect is proportional to the gradient of the axion field, as seen above.  For a lab experiment searching for dark matter or dark energy this effect would usually be suppressed by the low velocity of the axion field ($\sim 10^{-3}$ for the relative velocity between the earth and either the dark matter or dark energy).  However if we boost the precessing particle up to relativistic velocities then the effect is significantly increased.  While this is of course not possible for most sensitive spin precession experiments, there is one type of sensitive experiment with relativistic spins: a storage ring experiment such as a muon g-2 or proton EDM measurement (see e.g.~\cite{Anastassopoulos:2015ura}).  As we will see, the proton storage ring EDM proposal can indeed be used to search for axion dark matter and dark energy.

\begin{figure}
\begin{center}
\includegraphics[width=5.75 in]{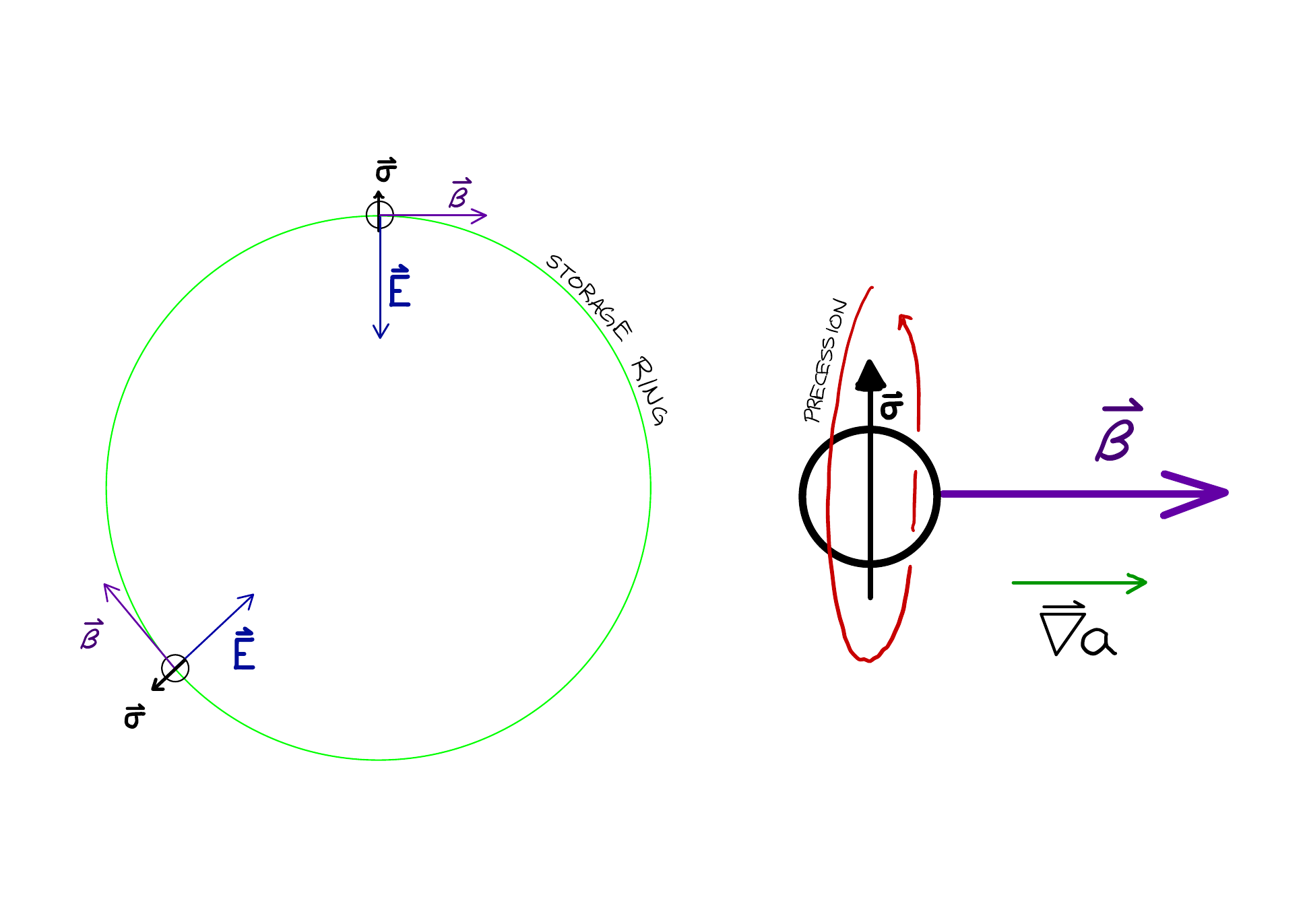}
\caption{ \label{Fig:storage ring schematic}  A sketch of the geometry for this storage ring proposal (left figure) and the directions of the proton's spin $\vec{\sigma}$, velocity $\vec{\beta}$ and precession, as well as the axion field gradient seen by the proton (right figure).  The proton's spin must be oriented radially and will then precess around its velocity (out of the plane of the ring).}
\end{center}
\end{figure}

When a proton is boosted to relativistic speeds it sees a much larger spatial gradient of the axion field directed along the direction of the boost.  The proton's spin will then precess around this direction.  Of course in a storage ring the velocity sweeps in a circle and so the net precession would generally cancel out (or be significantly reduced) as the proton goes around the ring many times.  To fix this we can use the `frozen spin' method where the spin of the proton is always locked to a fixed angle with respect to the velocity (see Fig.~\ref{Fig:storage ring schematic}).  Then the effect will add up over the entire orbit of the proton around the ring.  In this way we will boost the signal and be able to add it up over the entire integration time of the storage in the ring ($\sim 1000 \, \text{s}$ for the proton EDM experiment).

The proton storage ring EDM proposal uses this `frozen spin' method \cite{Anastassopoulos:2015ura, Haciomeroglu:2018nre}.  In that proposal the proton is placed in a ring with a large electric field (either all electric or a hybrid electric-magnetic design) and its spin is aligned with its velocity.  The proton is given the `magic momentum' so that the ring's electromagnetic fields cause the spin to precess by $2 \pi$ in exactly the time the proton orbits the ring once, thus keeping the spin and velocity always aligned.  This can be seen easily in the proton's rest frame where the large radial electric field looks like it has a large magnetic component perpendicular to the plane of the ring.  If the proton has an EDM, then its spin will also precess around the large electric field and thus out of the plane of the ring.  The protons spins are measured continuously over the period of about $1000 \, \text{s}$ that they spend in the ring.

Our proposal is to use the same storage ring to search for time-varying dark energy (with axionic couplings) and axion (or vector) dark matter.
Note that for the axion case, as in Fig.~\ref{Fig:storage ring schematic}, the proton's spin must be oriented radially (instead of tangentially as in the EDM case) so that it will precess around the proton's velocity, out of the plane of the ring.  The signal of axion dark matter or dark energy then is a small rising component of the proton's spin out of the plane as a function of time.  
Thus the same storage ring can be used to search for dark matter and dark energy as will be used for the proton EDM.  We are just searching for a different signal with a different dependence on the spin orientation of the proton and also, in the case of dark matter, with a fixed temporal frequency (see Section \ref{sec:theory}).

\begin{figure}
\begin{center}
\includegraphics[width=6 in]{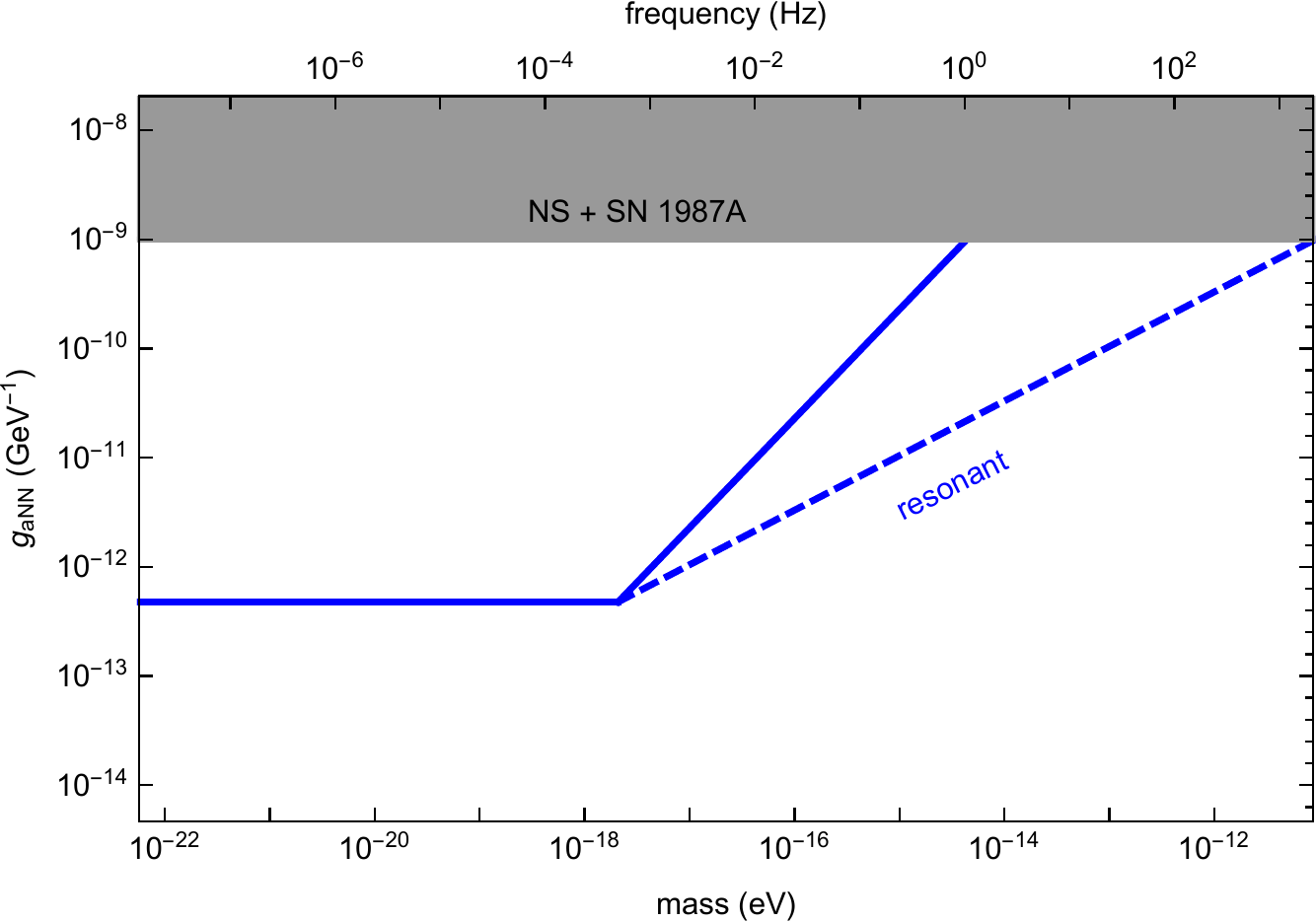}
\caption{ \label{Fig:axion sensitivity} The sensitivity of the storage ring proposal to axion dark matter in axion-nucleon coupling $g_\text{aNN}$  vs mass of axion.  The gray region is excluded by excess cooling in neutron stars and SN1987A.  The solid blue line shows the estimated sensitivity of the proton storage ring experiment assuming a $1000 \, \text{s}$ storage time and sensitivity to a proton spin precession rate of $10^{-9} \, \text{rad/s}$.  The dashed blue line shows the sensitivity of a resonant version of the experiment.}
\end{center}
\end{figure}

\begin{figure}
\begin{center}
\includegraphics[width=6 in]{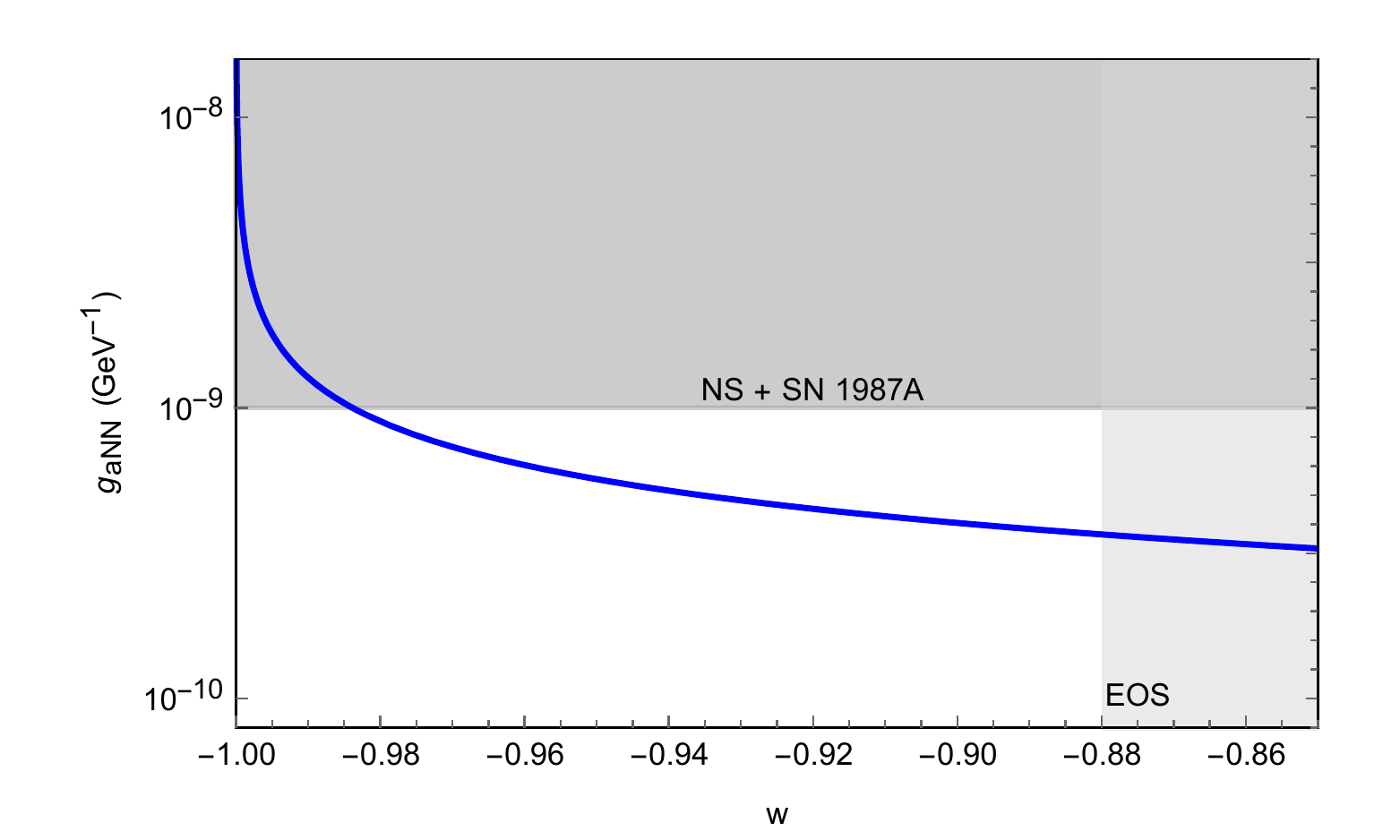}
\caption{ \label{Fig:dark energy sensitivity} The sensitivity of this storage ring experiment to time-varying dark energy with an axion-nucleon coupling $g_\text{aNN}$  vs equation of state of dark energy $w$.  The gray region is excluded by excess cooling in neutron stars and SN1987A.  The solid blue line shows the projected sensitivity of the proton storage ring experiment assuming a 1000 s storage time and sensitivity to a proton spin precession rate of $10^{-9} \, \text{rad/s}$.}
\end{center}
\end{figure}

In Figure \ref{Fig:axion sensitivity} we show an estimate for the sensitivity of this proposal to axion dark matter.  Figure \ref{Fig:dark energy sensitivity} shows the sensitivity to time-varying dark energy, assuming it has an axion-like coupling.  We have assumed similar numbers to the storage ring EDM proposal, namely the spin coherence time of the proton beam in the ring is $1000 \, \text{s}$ and that overall the sensitivity of the entire experiment after all protons have been sent through the ring will allow signals as small as a proton spin precession rate of $10^{-9} \, \text{rad/s}$ out of the plane of the ring to be measured.  For simplicity we have enveloped the sensitivity curve to remove the spikes that come from having a fixed time of $1000 \, \text{s}$.  This could in practice be achieved by varying this time by an order one factor from shot to shot.

The current astrophysical bounds on the axion-nucleon (really we consider only the axion-proton) coupling come from excess cooling of supernovae and neutron stars \cite{Chang:2018rso, PDG2019, Hamaguchi:2018oqw, Sedrakian:2015krq}.  While there is some uncertainty in these due to modeling of the astrophysical object, these bounds can not move too far and we have quoted an average value.

For axion dark matter frequencies above $\sim (1000 \, \text{s})^{-1}$ this experiment loses sensitivity because the axion signal averages out over the time each proton spend in the ring.  We could gain back some of this sensitivity by doing a resonant search.  If the proton momentum is tuned slightly away from the magic momentum then the spin will precess at a rate slightly different than the rate at which velocity is rotating.  The difference in the spin precession frequency and the frequency with which the velocity changes direction (which is the orbital frequency) is then the resonant frequency of this experiment.  By changing the detuning from the magic momentum, the resonant frequency can be swept to search for the axion.  The sensitivity of this resonant technique is shown as the dashed line in Fig.~\ref{Fig:axion sensitivity}.

\begin{figure}
\begin{center}
\includegraphics[width=6 in]{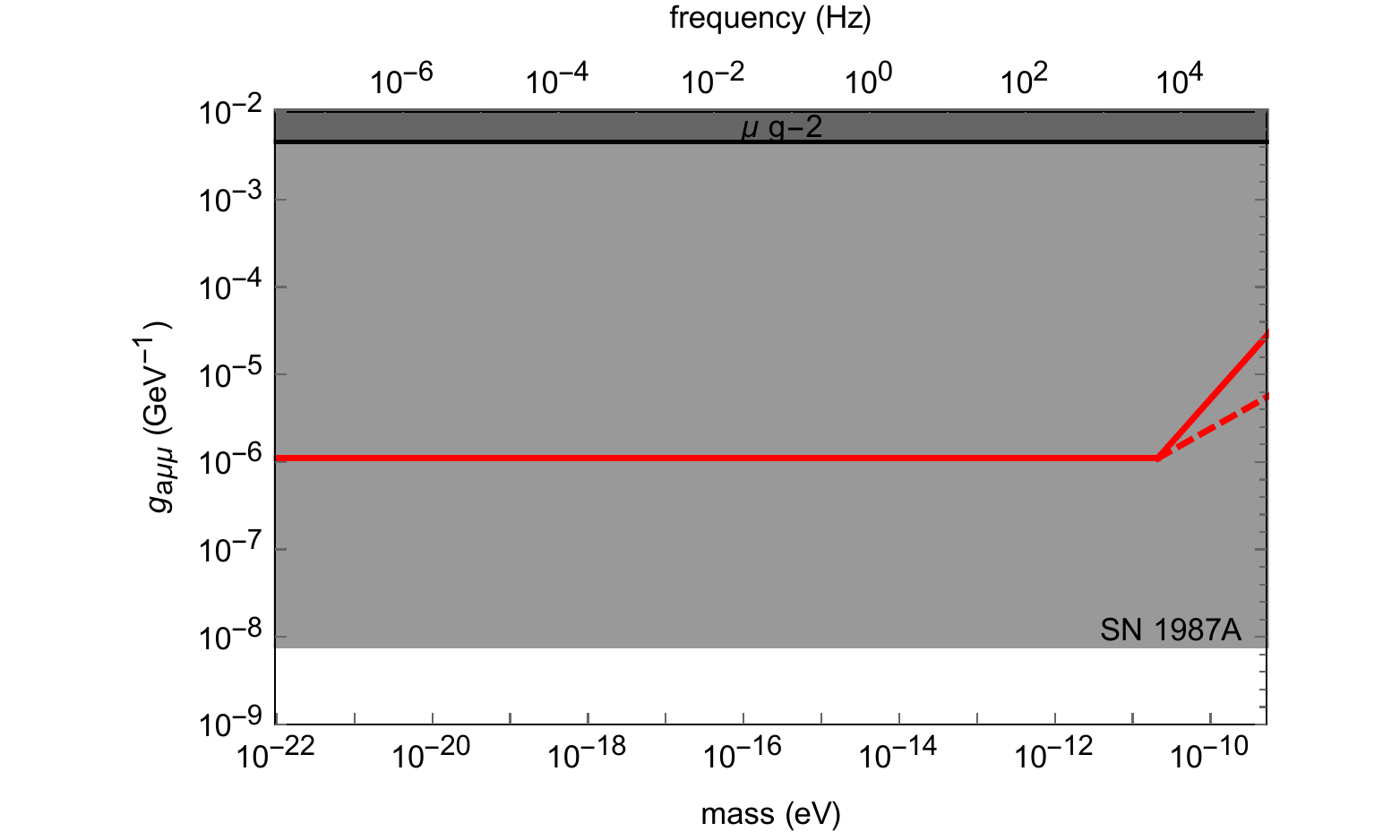}
\caption{ \label{Fig:muon DM sensitivity} The sensitivity of the storage ring proposal (with muons) to axion dark matter with an axion-muon coupling $g_{\text{a} \mu \mu}$  vs mass of axion.  The gray region is excluded by excess cooling in SN1987A \cite{Bollig:2020xdr}.  The dark gray region is the best current laboratory bound coming from virtual axion contributions to the muon g-2 \cite{Andreas:2010ms}.  The solid red line shows the estimated sensitivity of the storage ring experiment with muons assuming with parameters as explained in the text.  The dashed red line shows the sensitivity of a resonant version of the experiment.  The current Muon g-2 experiment at Fermilab would have a sensitivity a little over an order of magnitude worse than the red line over most of frequency space.}
\end{center}
\end{figure}

The Muon g-2 experiment at Fermilab is also a storage ring experiment and could in principle be used to search for axion dark matter with a coupling to muons, though since the spin is not frozen the sensitivity would be dominantly only in one narrow frequency band.  Muons could also in principle be placed in the same proton EDM storage ring.  In this case the muons' spins could be frozen and we could have sensitivity to a wide range of dark matter frequencies (and also to dark energy).  We show the estimated sensitivity for such a dark matter search in Fig.~\ref{Fig:muon DM sensitivity}.  As example parameters for such an experiment with muons we have assumed a total sensitivity to an axion-induced muon precession rate of $2.3 \,\frac{\text{mrad}}{\text{s}}$ in the ring.  Since the Fermilab g-2 experiment does not freeze the muons' spins relative to their velocity, the total sensitivity of that experiment ends up being a little over a factor of 10 worse than the dedicated experiment.  Of course the Fermilab experiment currently has data and could do such an analysis now which would already give the best laboratory bound on the axion-muon coupling.

The strongest current limit on the axion-muon coupling comes from SN1987A \cite{Bollig:2020xdr}.
There are also bounds from the number of degrees of freedom produced in the early universe, $N_\text{eff}$ \cite{Brust:2013xpv, DEramo:2018vss} which are not currently as constraining but will improve significantly with CMB S4 \cite{Baumann:2016wac} (and see approximate CMB bound in \cite{Bollig:2020xdr}).
The strongest laboratory bound comes from the virtual axion contributing to the muon g-2, which we quote as the ``$\mu$ g-2'' region in Figure \ref{Fig:muon DM sensitivity} \cite{Andreas:2010ms}.

Finally, we note that for vector dark matter and the magnetic dipole operator, Figure \ref{Fig:axion sensitivity} applies replacing $g_{\rm aNN} \rightarrow g'_{\rm MDM}/2$. the factor of 2 comes from a combination of normalization as well as averaging over the $\cos\phi$, the angle between the dark $E$-field and the normal to the ring (though the daily modulation of this signal adds an interesting effect).

The electric dipole operator could be effectively tested if the spin of the orbiting particle were frozen in the lab frame.  Such a set up would be possible with a particle like the deuteron, which has a negative anomalous magnetic moment.  The change in the particle spin in the lab frame is
\begin{equation}
    \frac{d\vec{\sigma}}{dt} = \frac{e}{m} \vec{\sigma} \times \left[ \left(a + {1\over\gamma}\right) \vec{B} - \frac{a \gamma}{\gamma + 1} \vec{\beta}(\vec{\beta}\cdot\vec{B}) - \left(a + \frac{1}{\gamma + 1}\right)\vec{\beta}\times\vec{E}
    \right]
\end{equation}
where $e/m$ is the charge to mass ratio and $a=(g-2)/2$ is the anomalous magnetic moment.   We see for a storage ring with a radial electric field, the spin stays fixed if $\gamma=- (1/a) - 1$.  For a ring with a perpendicular magnetic field, we require $\gamma = -(1/a)$.  For deuterium ($a\simeq -.14)$, these gamma factors are roughly 6 and 7 respectively.

\section{Experimental setup and backgrounds overview}
\label{sec:Experimental setup and backgrounds overview}

The experiment in most part is described in \cite{Anastassopoulos:2015ura}  where an all-electric ring is assumed, with some very important differences shown here.  In that case, the bending is provided by a pair of vertical plates in cylindrical shape to accommodate the particle bending horizontally and the focusing is provided by  electro-static quadrupoles.  The main challenge in this experiment is its sensitivity to the unwanted magnetic fields that inevitably would be present, requiring novel methods of detecting them.  The technology to accomplish the required level of cancellation has been developed.\cite{Haciomeroglu:2018bpm}  The hybrid version~\cite{Haciomeroglu:2019gp} of the method, however, reduces it automatically by several orders of magnitude via alternate magnetic, instead of electric focusing.  In general, the ring main focusing method also determines the nature of the prominent systematic error.  In rings with electric focusing, the main systematic error is unwanted magnetic fields, while in rings with magnetic focusing the main systematic error is an out of plane electric field.  When the effect of electric forces on beam dynamics is balanced by magnetic force or vice versa, an unwanted vertical spin precession is created. In addition, the compensated gravitational forces also induce a significant vertical spin precession that can only be resolved by CW and CCW storage.~\cite{Orlov:2012gr} A stored particle is assured to feel a total vertical force equal to zero by definition, but the delicate balance of those forces is critical.  The effect of those forces on the vertical spin precession rate depends on the particle's velocity and spin direction and in general it is different than when they balance each other in beam dynamics.  In general, they don't balance in spin dynamics nor by subtracting the effects from clockwise (CW) and counter-clockwise (CCW) beams, and thus they present a serious systematic error source.  Even though the gravity effect cancels comparing the CW and CCW effects, and in practice it is not an issue, the balancing act of combined electric and magnetic fields is more complicated and presents a serious challenge.
Both focusing cases listed above are very challenging to achieve; either the required cancellation of unwanted magnetic fields or the alignment of the electric fields in the bending sections.  However, in the hybrid version where alternate magnetic focusing is applied, it is also possible to store beams simultaneously traveling CW and CCW, which exactly cancels all out-of-plane dipole electric fields; the main systematic error source in this case.  Cancellation of higher than dipole order electric fields require CW-CCW beam closed orbits to trace the same path around the ring to 0.01mm. Therefore, the claim is that the hybrid lattice ring is in general easier and cheaper to achieve and this is what we will assume for the rest of the document.

\begin{figure}
\begin{center}
\includegraphics[width=5 in]{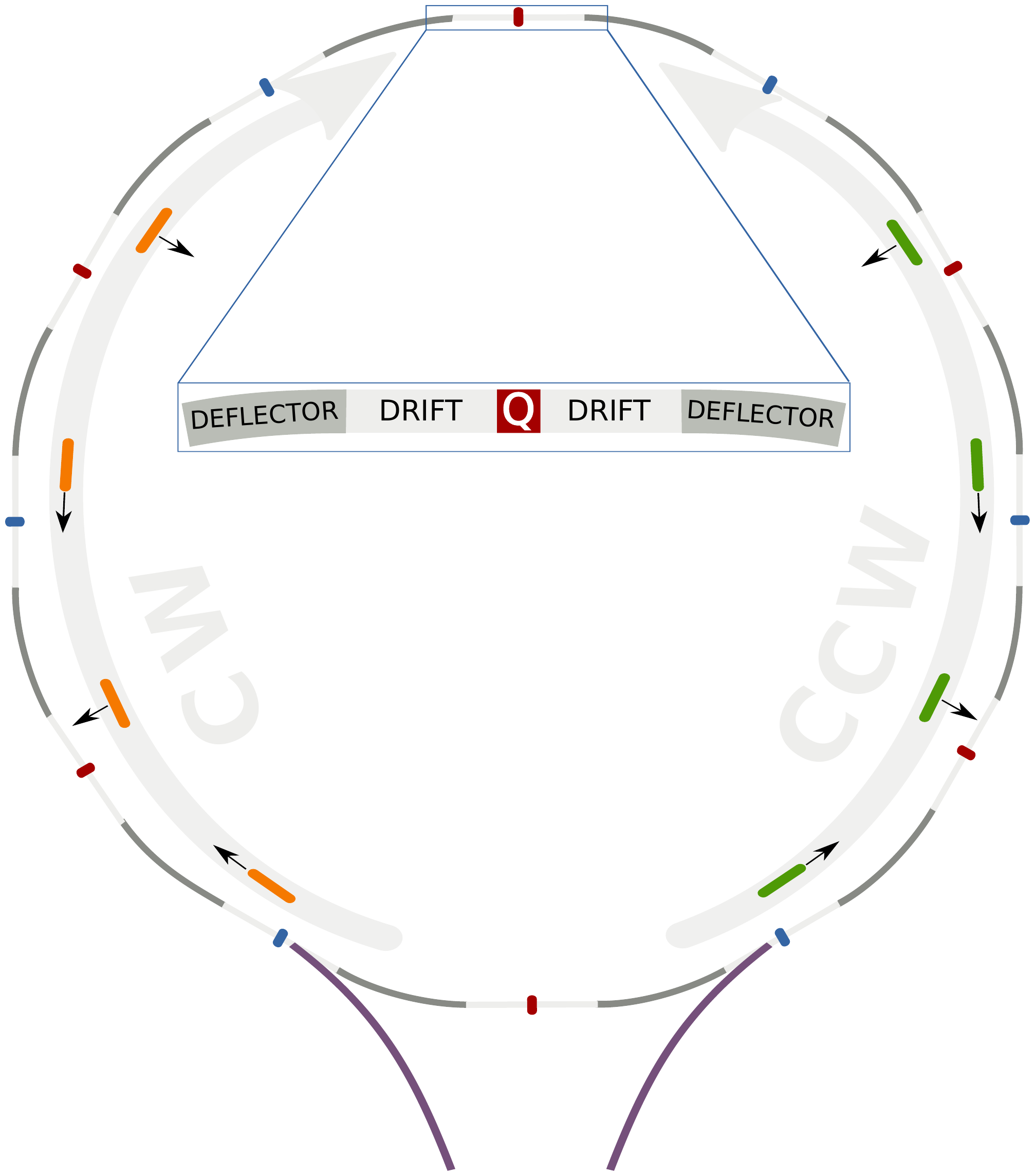}
\caption{ \label{Fig:alternating_symmetric_ring} The ring lattice is symmetric for CW and CCW beams, but also with respect to each lattice element in the ring. The ring lattice parameters are given in Table~\ref{tab:lattice}.  This design has been shown to be more robust against lattice-element position tolerances since the particle deviations from its ideal orbit cancel to high degree, when integrated around the ring.  The beam bunch polarization is also indicated, showing that all four spin directions are simultaneously stored for both CW and CCW beams for systematic error cancellations.}
\end{center}
\end{figure}

In the storage ring proton EDM experiment the beam is  bunched by a radio-frequency cavity (RF-cavity) inducing the so-called synchrotron oscillations~\cite{Benati:2012,Benati:2012err} for the captured protons.  Within the bunch, the proton momentum oscillates with an amplitude  depending on its momentum deviation from the average momentum and its relative arrival time at injection.  The RF-cavity is designed to equalize, to first order, all the proton momenta to the desired momentum value.  In one such momentum value, the so-called ``magic'' momentum, given by $p = mc / \sqrt{G} \approx 0.7$GeV/c for the proton, 
where $m$ is the proton mass, $c$ is the speed of light and $G=(g_p-2)/2=1.792847356$ the proton anomaly, the relative angle in the horizontal plane between the proton spin and momentum vectors will remain ``frozen'' if the horizontal bending is caused by only electric fields.~\cite{Anastassopoulos:2015ura}  There are two main reasons for applying the RF-cavity in the proton EDM experiment, it increases the spin coherence time of beams, e.g., with $dp/p \approx 10^{-4}$ from the milliseconds range without it to several seconds.~\cite{Orlov:2015,Orlov:2015_2, Anastassopoulos:2015ura} In addition,  it is essential in eliminating the polarimeter systematic errors~\cite{Brantjes:2012} arising from a  potential beam motion in position and angle relative to the external target and detector system. The spin coherence time can be further increased by the addition of sextupoles at an appropriate place~\cite{Orlov:2015,Orlov:2015_2, Anastassopoulos:2015ura, Guidoboni:2016} in the ring and/or by reducing the captured beam phase-space at injection.  A single RF cavity can capture both CW and CCW traveling particles.\footnote{The RF-cavity operates in the $\rm TM_{010}$ mode with the electric field in the longitudinal direction and a resonant frequency a multiple of the proton revolution frequency. In a regular storage ring, only half the RF cycle captures particles in stable synchrotron oscillations. The counter-rotating beams are captured in the opposite cycles matching the direction of motion.}  The number of bunches captured in the EDM ring is of order of $10^2$, while the polarization direction in half the bunches is kept in the longitudinal direction and of those, half of them with positive and the other half with negative helicity.  The remaining half bunches have the proton spin oriented in the radial direction half of which again are pointing radially outward and half inward,\footnote{It is also possible to store beams with vertical polarizations, half of them directed upwards and the other half downwards as long as there are stored bunches with horizontal polarizations within the same storage time to indicate that the horizontal spin component is ``frozen'' relative to the momentum vector.  This option can be explored to probe the EDM and the DM/DE cases simultaneously and to study additional systematic error sources.} similarly to those shown in Fig.~\ref{Fig:alternating_symmetric_ring}.  

\begin{figure}
\begin{center}
\includegraphics[width=5 in]{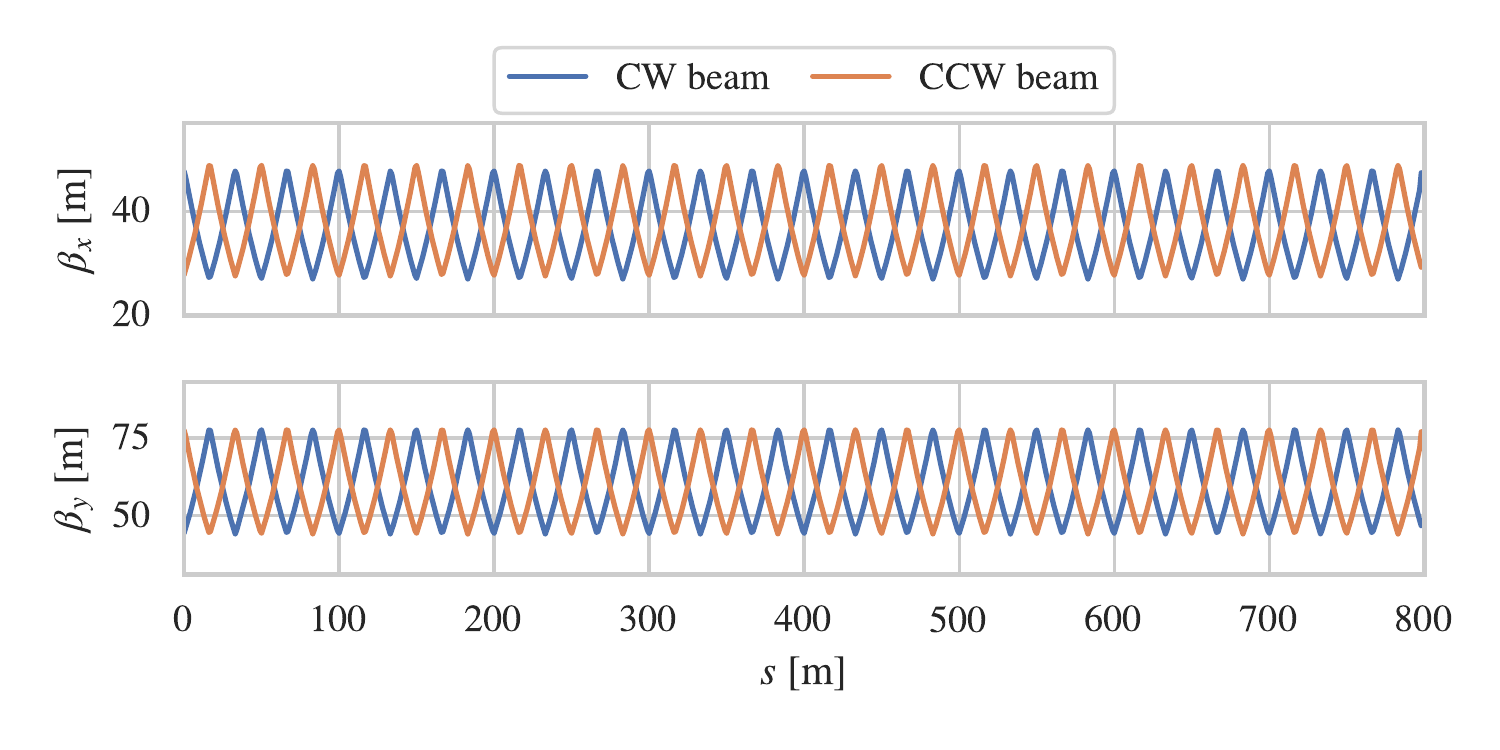}
\caption{ \label{Fig:beta_function_only} The horizontal and vertical beta-functions as a function of the longitudinal position around the ring. The two features, highly symmetric beta functions along the ring azimuth and quad polarity flipping, significantly supress the EDM and DM/DE systematics errors.  While the beam envelop oscillates going around the ring, the center of the quads remain ideally at zero for all of them.}
\end{center}
\end{figure}

\begin{figure}
\begin{center}
\includegraphics[width=5 in]{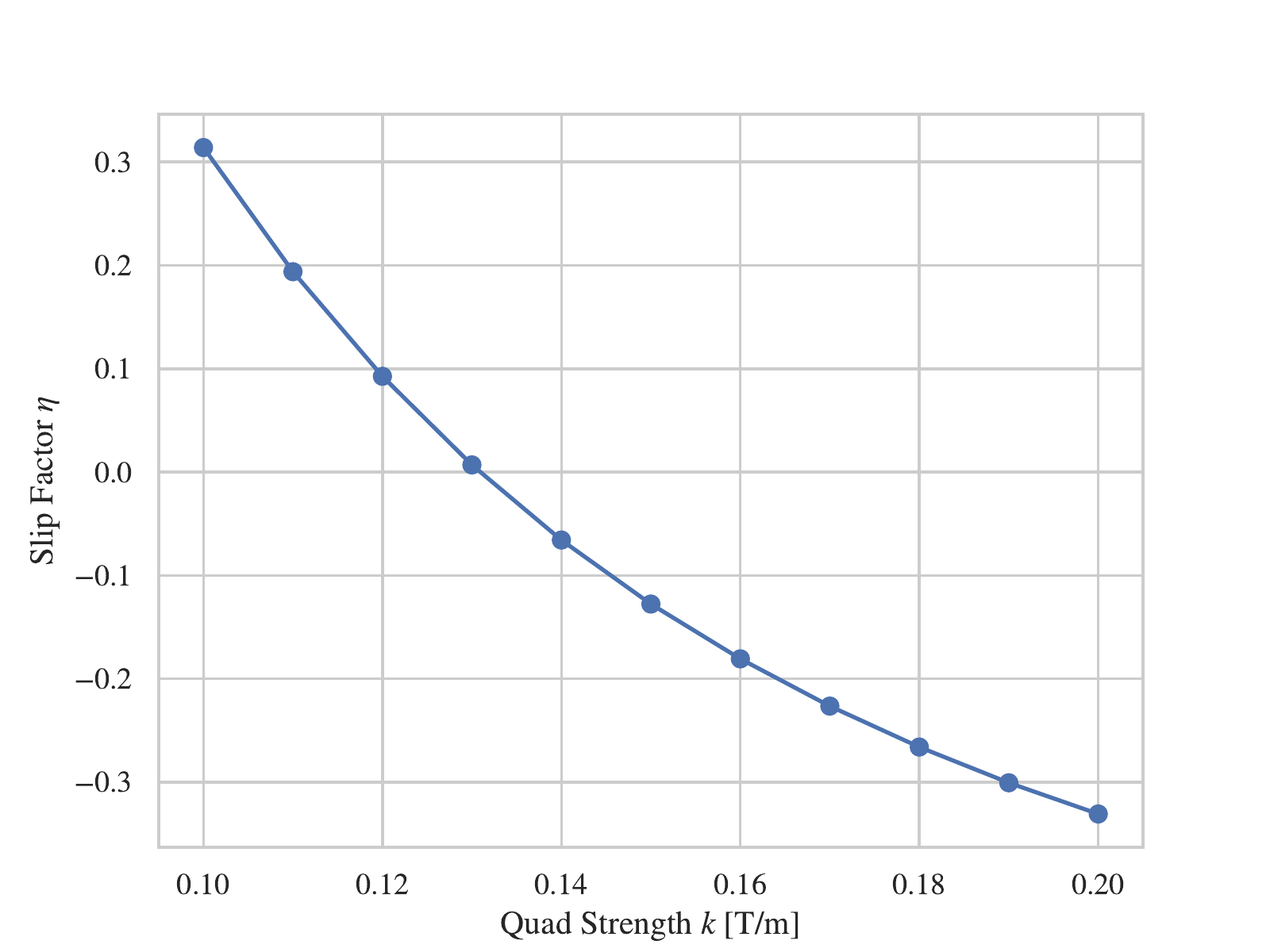}
\caption{ \label{Fig:slip_factor} The lattice slip factor as a function of the quadrupole strength.  The negative slip factor puts the ring lattice below transition, an essential quality in having low intra-beam-scattering and a stable storage for high beam intensities as required by the experiment.~\cite{Anastassopoulos:2015ura}}
\end{center}
\end{figure}

The longitudinal polarizations probe the particle EDM, while the radial polarizations probe the axion dark matter and/or dark energy of vacuum (DM/DE).  The positive and negative helicities are mainly used to eliminate the polarimeter systematic errors, while the CW and CCW injections are used to eliminate systematic errors related to the background (unwanted) electromagnetic fields.  Fig.~\ref{Fig:beta_function_only} shows the horizontal and vertical beta functions, indicating a highly symmetric beam envelop, a critical feature to reduce the sensitivity of the experiment on the lattice elements positioning errors.   The beta-function shows the stored beam envelop for each direction, which flips as the current polarity on the magnetic quadrupoles is flipped, providing one more tool combating the systematic errors. Fig.~\ref{Fig:slip_factor} shows the slip factor as a function of the quadrupole strength.  The negative slip-factor is required to establish equilibrium~\cite{Anastassopoulos:2015ura} between the three phase-space components of the stored beam due to intra-beam-scattering (IBS), potentially allowing to increase the beam storage lifetime to more than $10^4$~s.

We have used high-precision simulation software with the required accuracy, similar or better than,~\cite{Metodiev:2015ana}. Field imperfections and lattice misalignments were included for the main systematic error assessments (More details are given in Ref.~\cite{Omarov:arxiv}. The interface between the fields and the straight section is a hard-edge approximation, with no fringe fields.  The realistic fringe-fields of infinitely-high electric field plates with cylindrical geometry, as studied in~\cite{Metodiev:2014fri}, do not have a significant effect on particle beam/spin dynamics.  The beam parameters of the ring lattice used here are given in Table~\ref{tab:lattice}, with a main characteristic that of the high symmetry of the lattice elements.

\begin{table}[ht!]
\caption{\label{tab:lattice}The lattice parameters for the storage-ring proton EDM experiment.}
\begin{tabular}{l l l}
\hline
Parameter & Magnitude & Description \\
\hline
$p_0$ & 0.71 GeV/$c$ & Magic momentum \\
$\beta$ & 0.59 & $=v/c$, the particle speed \\
$R_0$ & 95.5 m &Deflector radius\\
$C$ & 800 m & Ring circumference \\
$f_c$ & 0.22 MHz & Cyclotron frequency\\
$f_x$ & 0.51 MHz & Horizontal betatron frequency \\
$Q_x$ & 2.3 & Horizontal betatron tune \\
$f_y$ & 0.49 MHz & Vertical betatron frequency \\
$Q_y$ & 2.2 & Vertical betatron tune \\
$E_0$ & 4.4 MV/m & Deflector electric field \\
$k$ & 0.2 T/m & Quadrupole strength \\
$L_{quad}$ & 40 cm & Quadrupole length \\
$L_{str}$ & 4.6 m & Straight section length (incl. quad.)\\
$N$ & 48 & Number of cells\\
\hline
\end{tabular}
\end{table}

\subsection{Backgrounds}
\label{subsec:backgrounds}

Our highly symmetric, hybrid ring lattice significantly relaxes the magnetic field shielding requirements as well as the magnetic quadrupole position tolerances.  In this section we are presenting the systematic errors and how we plan to eliminate them.  Depending on which direction the spin is frozen, which also determines the physics the beam is sensitive to, the systematic errors are not necessarily the same.  Running the experiment, with simultaneous storage of the bunched beams in all four different options, in both CW and CCW directions makes the experiment plausible. 
Table~\ref{tab:syserr1} shows the systematic errors for the EDM case and the remediation measures, while Table~\ref{tab:syserr2} shows the same for the radial polarization case.
We will address the two cases separately below. The main systematic error source is the large E-field bending sections: either a vertical offset of the counter-rotating beams (for the EDM case) or different vertical-angles of the counter-rotating beams (DM/DE case).

The effect of the external magnetic field is effectively shielded by the magnetic quadrupoles.  When magnetic focusing is used,  the vertical E-field is a major potential systematic error~\cite{deuteron_edm_proposal:2008deu}, since in its own rest frame it is partially converted into a radial magnetic field precessing the spin vertically.  However, the virtue of alternate magnetic focusing is that it also allows for simultaneous CW and CCW storage completely canceling the dipole vertical electric field effect. Vertical dipole E field originating somewhere in the ring is canceled by a single trim dipole E-field placed anywhere along the beam orbit for both CW and CCW directions. However, if the vertical E-field is not uniform and the counter-rotating beams do not map-out the same field on average, that will create a potential systematic error source, which can also be the dominant one in this case.

In the radial polarization case, the main systematic error source is the average vertical velocity vector, integrated inside all the electric  bending field sections.  In this case, the vertical velocity in a radial E-field location creates a rest-frame longitudinal B-field, which will rotate the spin into the vertical direction and it is similar to the DM/DE effect for the CW and CCW directions.  Therefore, any source of this systematic error needs to be symmetric enough so that it cancels when integrated around the ring within a single direction.  In a highly symmetric ring lattice, like the one assumed here, the distortion of the closed orbit, caused by the position error of the magnetic quads and of the bending sections, is symmetric and will cancel to high degree when integrating around the ring within a single direction.
On the other hand, any longitudinal magnetic field, caused by a possible non-zero current going through the ring in the vertical direction, would also cancel. The CW and CCW beams will have their spins rotated in the same direction as opposed to the DM/DE effect.

\begingroup
\begin{table}[h!]
\caption{\label{tab:syserr1}Main systematic errors and their remediation when hybrid fields (electric bending and magnetic focusing) are used for the EDM (longitudinal polarization) case. S-BPM: are the SQUID-based beam position monitors, see below.}
\begin{tabular}{ l  |   p{7.5cm} |}
{\bf Effect} & {\bf Remediation}  \\ \hline
Radial B-field. &  Magnetic focusing. \\ \hline
Unwanted vertical forces when other than  &  Vary the magnetic focusing strength and fit  for \\
 magnetic focusing is present. & the DC offset in the vertical precession rate.\cite{Haciomeroglu:2018nre} \\ \hline
Dipole vertical E-fields. &  Cancel exactly with CW and CCW beam storage. \\ \hline
Quadrupole E-field in the electric bending &  Probe it by locally splitting the counter-rotating \\ sections. & beams and cancel it with trim E-fields. Finally, keep the counter-rotating beams at the same position to S-BPM resolution.\\ \hline
Corrugated (non-planar) orbit.  & Minimize effect with symmetric lattice design. Finally, keep the CR beams at same position, at the electric field bending sections, using beam-based alignment. \\  \hline
Longitudinal B-field.  &  Small effect.   \\ \hline
Geometrical phase effect due to lattice  &  Equivalent to a spin resonance due to lattice \\
elements imperfections. & elements imperfections.  Magnetic quadrupoles: beam-based alignment to 1$\mu$m rms. E-field sections: Absolute beam position monitors to $<$0.1mm.   \\ \hline
Geometrical phase effect due to external  &  Equivalent to a spin resonance due to external \\ magnetic fields. & magnetic interference coupled with electric field bending section misplacement.\cite{Haciomeroglu:2019gp,Haciomeroglu:2017gp}  When the local spin effects are kept below 1nT B-field equivalent, the effect is negligible even for one directional (CW or CCW only) storage.\\ \hline
RF cavity vertical and horizontal misalignment. & Vary the longitudinal lattice impedance to probe the effect of the cavity's vertical and horizontal angular misalignments. The vertical and horizontal offsets are much smaller effects.\\  \hline
\end{tabular}
\end{table}
\endgroup

\begingroup
\begin{table}[h!]
\caption{\label{tab:syserr2}Main systematic errors and their remediation when hybrid fields (electric bending and magnetic focusing) are used for the DM/DE, probing (pseudo-)scalar fields, (radial polarization) case.}
\begin{tabular}{ l  |   p{7.5cm} |}
{\bf Effect} & {\bf Remediation}  \\ \hline
Radial B-field. &  Small effect. \\ \hline
Unwanted vertical forces when other than  &  Small effect. \\
 magnetic focusing is present. &   \\ \hline
Dipole vertical E-fields. &  Small effect. \\ \hline
Quadrupole E-field in the electric bending &  Small effect.\\ 
sections. & \\ \hline
Corrugated (non-planar) orbit.  & Minimize effect with symmetric lattice design. Finally, keep the stored beams at zero average vertical angle when integrating over the electric field bending sections. \\  \hline
Longitudinal B-field.  &   The CW and CCW stored proton spins rotate in {\it same} direction, while the (pseudo-)scalar fields rotate them in opposite directions.\\ \hline
Geometrical phase effect due to lattice  &  Equivalent to a spin resonance due to lattice \\
elements imperfections. & elements imperfections.  Magnetic quadrupoles: beam-based alignment to 1$\mu$m rms. E-field sections: Absolute beam position monitors to $<$0.01mm per injection.   \\ \hline
Geometrical phase effect due to external   &  Equivalent to a spin resonance due to external \\
magnetic fields. & magnetic interference coupled with electric field bending section misplacement.\cite{Haciomeroglu:2019gp,Haciomeroglu:2017gp}  When the local spin effects are kept below 1nT B-field equivalent, the effect is negligible even for one directional (CW or CCW only) storage. In this polarization case, the relevant fields and lattice misplacements may be in a different direction than the previous table.  \\ \hline
RF cavity vertical and horizontal misalignment. & Small effect.  \\  \hline
\end{tabular}
\end{table}
\endgroup

\subsection{Ring lattice design}

The tools we have in our disposal to minimize and eliminate the potential systematic errors are:

\begin{enumerate}
\item Design a symmetric lattice to minimize the effect of the lattice-elements positioning errors.
\item Shield the external magnetic field anywhere in the storage ring to below 100~nT.  This can be easily achieved by using an array of fluxgate magnetometers and shimming the residual field using an array of current wires.  The aim for the low azimuthal harmonics, i.e. $N=1, 2, ...,10$, of the B-field is to reach below 1~nT amplitude.
\item The SQUID-based beam position monitors (S-BPM) developed at IBS-CAPP in collaboration with KRISS in Korea utilizing special, low-temperature-superconducting (LTS) SQUID-gradiometers, have been showed to exhibit 10~nm/$\sqrt{\rm Hz}$,\cite{Haciomeroglu:2018nre} i.e. every 100~s they can provide 1~nm beam separation resolution.   The S-BPMs will be used to reduce the external B-field and the magnetic quadrupole position tolerance by modulating the vertical tune and by using beam-based alignment.\cite{Tenenbaum:2000prs,Wagner:2018bba}
\end{enumerate}

\subsection{Position tolerances of ring elements}

Here, we will address the main systematic errors in both the EDM and DM/DE cases in more detail, their origin, and the way to eliminate them. The main issue in the EDM case arises from the subtle effect that there might be a quadrupole E-field present, most likely at the electric bending sections, while there is  a radial B-field present somewhere in the ring.  A radial B-field of any order will inevitably split the vertical position of the counter-rotating beams, which again will mean that the counter-rotating beams will sense vertical electric fields of opposite direction or at least not of the same amplitude and therefore, the cancellation will only be partial.  Only same direction/same amplitude vertical E-fields cancel with CW and CCW beams, meaning that the counter-rotating beams will cancel the vertical dipole electric field (uniform field), but not higher order when the beams do not overlap at the required precision.  The unknown in this case is the vertical E-field focusing index in the E-bending sections, which needs to be designed and installed as well as possible.  Technically, it is possible to mechanically align the plates enough to achieve an average of 1.1 field focusing index (requiring an average alignment of order $10 \, \mu$m between the two plates and a similar plate flatness over 20~cm vertically, integrated over a 3~m azimuthal length of the plates).  We plan to use a local radial B-field to separate the two counter-rotating beams by as much as 1~mm locally at each bending section, and apply a trim quadrupole E-field at the specific E-bending section to eliminate the vertical non-uniformity of E-fields to the needed accuracy.  We will be able to reduce the unwanted electric quadrupole component of the E-field sections to well below 1.001 level, so that a local separation of the counter-rotating beams of less than $0.1 \, \mu$m is adequate.  The issue is to keep the plate alignment stability to better than $0.1 \, \mu$m over $10^3$s, which should not be a major challenge.

Our S-BPMs have an estimated resolution of 10~nm/$\sqrt{\rm Hz}$,~\cite{Haciomeroglu:2018nre}, but even a resolution of $0.1 \, \mu$m within a fill of $10^3$s duration of the beam separation  between CW and CCW directions, would be adequate.
Assuming a standard deviation in the vertical electric field focusing index in the E-bending sections of 1.001 would be adequate to reduce the main systematic errors below the statistical sensitivity.
  The total experiment duration is assumed to be $4 \times 10^7$s, meaning approximately $4 \times 10^4$ fills, resulting to an unwanted maximum vertical spin precession rate of  $\approx 0.75$~nrad/s, corresponding to an EDM of $0.75 \times 10^{-29}\, e \cdot {\rm cm}$ in the same ring.      
A $10 \mu$m resolution in the position of the quadrupoles has been demonstrated using beam-based alignment in DC-mode in hadronic storage rings~\cite{Wagner:2018bba}, while the S-BPMs and the method we have developed to probe the average separation of the CR beams, have much better resolution than that.  

Next, we are addressing the radial polarization case. Here, as stated previously, it is the average vertical velocity, when integrated over all  electric bending sections, which dominates the systematic error source.   Systematic error studies with high-precision simulation indicate that a symmetric ring  with the magnetic quadrupoles position-error randomly distributed with a standard deviation of $1 \, \mu$m, and a quadrupole component in the electric field bending sections with a sigma of 1.001, the systematic error is below the aimed statistical error of $10^{-29}\, e \cdot {\rm cm}$.

Constructing and setting up the ring that can deliver the aimed sensitivity requires a thorough study of all possible systematic error sources.  Those listed in Tables~\ref{tab:syserr1},~\ref{tab:syserr2} are the main ones regarding the spin-related systematic errors, while the systematic errors related to the polarimeter detector are adequately addressed by simultaneously storing positive and negative helicity beams.\cite{Brantjes:2012}

\subsection{Construction, commission, and running the experiment}

The sequence of constructing, commissioning and running of the experiment is:

\begin{enumerate}

\item The ring lattice is symmetric, i.e., the whole ring is designed to be approximately 800~m long in circumference, consisting of 48 identical units of 16.67~m long each.  Each unit consists of a bending segment made of electric field plates 12~m long each and one magnetic quadrupole of 0.4~m long capable of delivering at least 0.2T/m magnetic field made completely of wires.

\item In addition, there is one  RF-cavity for beam bunching,  two beam-injection points with kickers (if they are magnetic they need to be made out of wires), and one RF-dipole to rotate the spin from vertical at injection to the required horizontal direction.  Furthermore, there are a couple of polarimeter locations equipped with the required detectors to observe the spin precession rate in the horizontal and vertical direction for both CW and CCW stored beams.  Other essential elements, e.g., beam-position-monitors (BPM) based on sensing either the beam electric field or the beam generated magnetic field are also present as needed.  There are SQUID-based BPMs (S-BPMs), just before and after each electric field bending section to probe the separation of the counter-rotating (CR) beams and possibly paired with button-BPMs for electric field sensing, to probe the transverse position of the CR-beams.

\item The electric field plates are separated by 3~cm of vertical, metallic plates of cylindrical geometry made as parallel as possible.  In the rest of the ring there are parallel thin plates shielding the proton beams from vertical electrical forces caused by the induced charges on the ground plates.  Since the beams are only moderately relativistic, we need to design the electric field plate geometry to minimize the lattice impedance and also to be able to vary it for systematic error studies.

\item The vacuum chamber should be made out of bake-able aluminum, capable of reaching low $10^{-10}$~Torr in order to minimize beam-gas scattering.  The aluminum and all materials near the beam should be of low magnetization (probably not-recycled aluminum) in order to keep the permanent-stray-magnetic-field in the storage region below 100~nT at all locations.

\item Construct the ring lattice with 1~mm position mechanical tolerance.  Vertically (critical), using water tubes, minimize the lattice corrugation to below $100 \, \mu$m, with a special attention to keep the low azimuthal harmonics to below $10 \, \mu$m amplitude.\cite{Baklakov:1998gmt,Shiltsev:2002gmt}

\item Install a current carrying wire network to be able to cancel unwanted external magnetic fields paired with fluxgate magnetometers with enough resolution to achieve below 1~nT for the low azimuthal harmonics of the magnetic field amplitude in all directions.

\item Inject into the ring both CW and CCW beams with vertical polarization direction (stable spin direction), and let the beams de-bunch without powering the RF-cavity.  Then, power the RF cavity slowly and re-bunch the beam into the required number of bunches.  

\item Apply the RF-solenoid to precess the spin from vertical into the four spin directions (forward, backwards, radially outwards and radially inwards) for both CR directions.  

\item Tune the RF-cavity frequency, so that the particle spins precess in the horizontal plane and measure the beam polarization value.  Next, freeze the spin  in the desired direction by using feedback signals from the polarimeters.  If a single RF-frequency doesn't freeze simultaneously the horizontal spin precessions of the CR beams, then apply a small vertical dipole magnetic field to freeze both of them.

\item The vertical alignment specs of the electric field plates are very strict, difficult to achieve mechanically.  However, in the hybrid ring and with simultaneous CW and CCW beam storage the effect is canceled in a straight forward way by placing a trim vertical dipole E-field anywhere in the ring.  The amplitude of the trim dipole E-field is adjusted by zeroing the vertical spin precession rate of either the CW or CCW going beam.

\item Apply beam-based alignment with a detector resolution better than $10 \, \mu$m, while paying attention to keep the low azimuthal harmonics, i.e. $N=1, 2, ...,10$, of the lattice elements deviation from the ideal location better than $0.1 \, \mu$m.

\item Minimize the separation of the CW and CCW beams before and after the electric bending sections, the main source of systematic errors.

\item Increase the lattice impedance to increase the sensitivity to the RF-cavity horizontal and vertical angle misalignments.  Rotate the RF-cavity to reduce the effect below the statistical sensitivity with one injection and then return the ring lattice impedance to minimum.

\item Increase the separation of the CR-beams to 1~mm  locally at one electric field bending section at a time.  This can be easily achieved by using radial magnetic fields before and after the electric bending sections.  This effect increases the sensitivity of the experiment to electric quadrupole component, which can be trimmed out to statistical sensitivity level.  Repeat for all 48 plate sections and then return to the minimum beam separation.

\item Run the experiment to probe the proton EDM and the DM/DE of vacuum with full sensitivity.

\end{enumerate}

Before the start of the ring construction, a test-section of one section of the ring will be constructed and tested to provide information regarding compatibility issues at the integration stage.  After that, the ring construction should take less than two years to accomplish, with the commissioning and systematic error studies taking less than six months to accomplish.  The experiment should take about four years of running to accumulate the required statistics for $10^{-29}\, e \cdot {\rm cm}$ sensitivity on the proton EDM and comparable sensitivity to the DM/DE.

\section{Discussion}
\label{sec:discussion}
Storage rings are competitive with atomic co-magnetometer spin precession experiments in searching for interactions between protons and cosmic pseudo-scalar and magnetic dipole vector backgrounds. This is because the relativistic  velocity between the protons and the background enhances the velocity dependent spin precession, compensating for the much larger shot noise in storage ring experiments in comparison to atomic co-magnetometer experiments. However, background vector fields that interact through electric dipole couplings are harder to probe using storage ring experiments. This is because the electric dipole precession is not amplified by the relativistic motion of the protons. Thus, due to the much higher shot noise in storage ring experiments in comparison to atomic co-magnetometer experiments, the latter will be significantly more sensitive to such vector interactions. Storage ring experiments to search for Lorentz violating cosmic backgrounds are thus complementary to atomic co-magnetometer searches for such backgrounds, allowing us to even distinguish the micro-physics of a signal in such experiments. 

It is also important to highlight that the search for cosmic backgrounds using storage ring experiments can use the same hardware as the setup used to search for the electric dipole moment of nucleons. The major difference in terms of the operation of the device concerns the orientation of the nucleon spin relative to the direction of motion of the proton around the ring. Implementation of this change should not be unduly onerous. While the backgrounds for an electric dipole moment search are different from those that would affect this cosmic search, the technologies developed to implement the electric dipole moment search can directly be applied to mitigate the corresponding backgrounds for the cosmic search. Thus, this laboratory search for dark energy and dark matter complements the use of storage rings to search for nucleon electric dipole moments.

\section*{Acknowledgments}
We would like to thank Savas Dimopoulos and Roni Harnik for discussions. P.W.G.~was supported by DOE Grant DE-SC0012012, by NSF Grant PHY-1720397, the Heising-Simons Foundation Grants 2015-037 and 2018-0765, DOE HEP QuantISED award \#100495, and the Gordon and Betty Moore Foundation Grant GBMF7946. S.R.~was supported in part by the NSF under grants PHY-1818899 and PHY-1638509, the Simons Foundation Award 378243 and the Heising-Simons Foundation grant 2015-038.  S.H., Z.O. and Y.K.S. were supported by IBS-R017-D1-2020-a00 of the Republic of Korea.


\begin{thebibliography}{10}
\expandafter\ifx\csname url\endcsname\relax
  \def\url#1{{\tt #1}}\fi
\expandafter\ifx\csname urlprefix\endcsname\relax\def\urlprefix{URL }\fi



\bibitem{Pospelov:2004fj} 
  M.~Pospelov and M.~Romalis,
  Phys.\ Today {\bf 57N7}, 40 (2004).
  doi:10.1063/1.1784301
  
  
  
\bibitem{Romalis:2013kkt} 
  M.~V.~Romalis and R.~R.~Caldwell,
  arXiv:1302.1579 [astro-ph.CO].
  
\bibitem{Bennett:2008cpt}
G.~W.~Bennett {\it et al.},
  Phys.\ Rev.\ Lett.\ {\bf 100}, 091602 (2008) 091602
  doi:10.1103/PhysRevLett.100.091602
  [arXiv:0709.4670 [hep-ex]].

  
\bibitem{Brown:2010dt} 
  J.~M.~Brown, S.~J.~Smullin, T.~W.~Kornack and M.~V.~Romalis,
  Phys.\ Rev.\ Lett.\  {\bf 105}, 151604 (2010)
  doi:10.1103/PhysRevLett.105.151604
  [arXiv:1006.5425 [physics.atom-ph]].


\bibitem{Graham:2017ivz} 
  P.~W.~Graham, D.~E.~Kaplan, J.~Mardon, S.~Rajendran, W.~A.~Terrano, L.~Trahms and T.~Wilkason,
  Phys.\ Rev.\ D {\bf 97}, no. 5, 055006 (2018)
  doi:10.1103/PhysRevD.97.055006
  [arXiv:1709.07852 [hep-ph]].
  
\bibitem{Gemmel:2010ft} 
  C.~Gemmel {\it et al.},
  Phys.\ Rev.\ D {\bf 82}, 111901 (2010)
  doi:10.1103/PhysRevD.82.111901
  [arXiv:1011.2143 [gr-qc]].
  
\bibitem{Graham:2013gfa} 
  P.~W.~Graham and S.~Rajendran,
  Phys.\ Rev.\ D {\bf 88}, 035023 (2013)
  doi:10.1103/PhysRevD.88.035023
  [arXiv:1306.6088 [hep-ph]].

\bibitem{Graham:2019bfu} 
  P.~W.~Graham, D.~E.~Kaplan and S.~Rajendran,
  Phys.\ Rev.\ D {\bf 100}, no. 1, 015048 (2019)
  doi:10.1103/PhysRevD.100.015048
  [arXiv:1902.06793 [hep-ph]].
  
\bibitem{Graham:2017hfr} 
  P.~W.~Graham, D.~E.~Kaplan and S.~Rajendran,
  Phys.\ Rev.\ D {\bf 97}, no. 4, 044003 (2018)
  doi:10.1103/PhysRevD.97.044003
  [arXiv:1709.01999 [hep-th]].
  
\bibitem{Budker:2013hfa} 
  D.~Budker, P.~W.~Graham, M.~Ledbetter, S.~Rajendran and A.~Sushkov,
  Phys.\ Rev.\ X {\bf 4}, no. 2, 021030 (2014)
  doi:10.1103/PhysRevX.4.021030
  [arXiv:1306.6089 [hep-ph]].

\bibitem{Farley:2003wt} 
  F.~J.~M.~Farley {\it et al.},
  Phys.\ Rev.\ Lett.\  {\bf 93}, 052001 (2004)
  doi:10.1103/PhysRevLett.93.052001
  [hep-ex/0307006].
  
\bibitem{Anastassopoulos:2015ura} 
  V.~Anastassopoulos {\it et al.},
  Rev.\ Sci.\ Instrum.\  {\bf 87}, no. 11, 115116 (2016)
  doi:10.1063/1.4967465
  [arXiv:1502.04317 [physics.acc-ph]].

\bibitem{Bailey:1979np}
J.~Bailey {\it et al.},
Nucl.\ Phys.\ {\bf B150}, 1-75 (1979).
doi:10.1016/0550-3213(79)90292-X

\bibitem{Haciomeroglu:2018nre} 
  S.~Hac\i\"omero\u glu and Y.~K.~Semertzidis,
  Phys.\ Rev.\ Accel.\ Beams {\bf 22}, no. 3, 034001 (2019)
  doi:10.1103/PhysRevAccelBeams.22.034001
  [arXiv:1806.09319 [physics.acc-ph]].

\bibitem{Chang:2018rso} 
  J.~H.~Chang, R.~Essig and S.~D.~McDermott,
  JHEP {\bf 1809}, 051 (2018)
  doi:10.1007/JHEP09(2018)051
  [arXiv:1803.00993 [hep-ph]].

\bibitem{PDG2019} 
M. Tanabashi et al. (Particle Data Group), Phys. Rev. D 98, 030001 (2018) and 2019 update.


\bibitem{Hamaguchi:2018oqw} 
  K.~Hamaguchi, N.~Nagata, K.~Yanagi and J.~Zheng,
  Phys.\ Rev.\ D {\bf 98}, no. 10, 103015 (2018)
  doi:10.1103/PhysRevD.98.103015
  [arXiv:1806.07151 [hep-ph]].
  
\bibitem{Sedrakian:2015krq} 
  A.~Sedrakian,
  Phys.\ Rev.\ D {\bf 93}, no. 6, 065044 (2016)
  doi:10.1103/PhysRevD.93.065044
  [arXiv:1512.07828 [astro-ph.HE]].

\bibitem{Bollig:2020xdr}
R.~Bollig, W.~DeRocco, P.~W.~Graham and H.~T.~Janka,
Phys. Rev. Lett. \textbf{125}, no.5, 051104 (2020)
[arXiv:2005.07141 [hep-ph]].

\bibitem{Brust:2013xpv} 
  C.~Brust, D.~E.~Kaplan and M.~T.~Walters,
  JHEP {\bf 1312}, 058 (2013)
  doi:10.1007/JHEP12(2013)058
  [arXiv:1303.5379 [hep-ph]].

\bibitem{DEramo:2018vss} 
  F.~D'Eramo, R.~Z.~Ferreira, A.~Notari and J.~L.~Bernal,
  JCAP {\bf 1811}, 014 (2018)
  doi:10.1088/1475-7516/2018/11/014
  [arXiv:1808.07430 [hep-ph]].

\bibitem{Baumann:2016wac} 
  D.~Baumann, D.~Green and B.~Wallisch,
  Phys.\ Rev.\ Lett.\  {\bf 117}, no. 17, 171301 (2016)
  doi:10.1103/PhysRevLett.117.171301
  [arXiv:1604.08614 [astro-ph.CO]].


\bibitem{Andreas:2010ms}
S.~Andreas, O.~Lebedev, S.~Ramos-Sanchez and A.~Ringwald,
JHEP \textbf{08}, 003 (2010)
doi:10.1007/JHEP08(2010)003
[arXiv:1005.3978 [hep-ph]].

\bibitem{Haciomeroglu:2018bpm} 
  S.~Hac\i\"omero\u glu, D.~Kawall, Yong-Ho Lee, A.~Matlashov, Z.~Omarov, and Y.~K.~Semertzidis,
  PoS ICHEP2018  (2019) 279
  doi:10.22323/1.340.0279 

\bibitem{Haciomeroglu:2019gp} 
  S.~Hac\i\"omero\u glu, Y.~Orlov, and Y.~K.~Semertzidis,
  Nucl.\ Instr.\ Meth.\  {\bf A927}, 262-266 (2019)
  doi:10.1016/j.nima.2019.01.046

\bibitem{Orlov:2012gr} 
  Y.~Orlov, E.~Flanagan, Y.~Semertzidis, 
  Phys.\ Lett.\ {\bf A376}, (2012) 2822-2829
  doi:10.1016/j.physleta.2012.08.011
 
\bibitem{Haciomeroglu:2017gp} 
  S.~Hac\i\"omero\u glu, and Y.~K.~Semertzidis,
  [arXiv:1709.01208 [physics.acc-ph]]


\bibitem{Benati:2012} 
  P.~Benati {\it et al.}, 
  Phys.\ Rev.\ ST\ Accel.Beams  {\bf 12}, 15 (2012) 124202
  doi:10.1103/PhysRevSTAB.16.124202
 
\bibitem{Benati:2012err} 
  P.~Benati {\it et al.}, 
  Phys.\ Rev.\ ST\ Accel.Beams  {\bf 12}, 16 (2013) 049901
  doi:10.1103/PhysRevSTAB.16.049901
 

\bibitem{Orlov:2015} 
  Y.~Orlov
  [arXiv:1504.07304 [physics.acc-ph]].


\bibitem{Orlov:2015_2} 
  Y.~Orlov
  [arXiv:1506.02069 [physics.acc-ph]].

\bibitem{Guidoboni:2016}
G.~Guidoboni {\it et al.}, 
  Phys.\ Rev.\ Lett.\  {\bf 117}, no. 5, 054801 (2016)
  doi:10.1103/PhysRevLett.117.054801
  
  
\bibitem{Brantjes:2012} 
  N.P.M.~Brantjes {\it et al.}, 
  Nucl.\ Instr.\ Meth.\  {\bf A664}, 49-64 (2012)
  doi:10.1016/j.nima.2011.09.055

  
\bibitem{Metodiev:2015ana} 
  E.M.~Metodiev {\it et al.}, 
  Nucl.\ Instr.\ Meth.\  {\bf A797}, 311-318 (2015)
  doi:10.1016/j.nima.2015.06.032
  [arXiv:1503:02247 [physics.acc-ph]]

\bibitem{Omarov:arxiv} 
  Z.~Omarov {\it et al.}, 
  [arXiv:2007.10332 [physics.acc-ph]]
  
  
\bibitem{Metodiev:2014fri} 
  E.M.~Metodiev {\it et al.}, 
 Phys.\ Rev.\ ST\ Accel.Beams  {\bf 17}, 7 (2014) 074002
  doi:10.1103/PhysRevSTAB.17.074002

\bibitem{deuteron_edm_proposal:2008deu}
 D.~Anastassopoulos {\it et al.},
  ``AGS Proposal: Search for a permanent electric dipole moment of the deuteron nucleus at the $10^{-29}\, e \cdot {\rm cm}$ level.,'' available from Brookhaven National Laboratory, 2008.
    
\bibitem{Tenenbaum:2000prs} 
  P.~Tenenbaum and T.O.~Raubenheimer 
 Phys.\ Rev.\ ST\ Accel.Beams  {\bf 3}, 5 (2000) 052801
  doi:10.1103/PhysRevSTAB.3.052801

\bibitem{Wagner:2018bba}
Tim Wagner {\it et al.},
Hyperfine Interact (2018) 239: 61
doi:10.1007/s10751-018-1539-6

\bibitem{Baklakov:1998gmt}
B.~Baklakov {\it et al.,}
 Phys.\ Rev.\ ST\ Accel.Beams  {\bf 1}, 3 (1998) 031001
  doi:10.1103/PhysRevSTAB.1.031001

\bibitem{Shiltsev:2002gmt}
V.~Shiltsev {\it et al.,}
available from \\
\url{icfa-nanobeam.web.cern.ch/icfa-nanobeam/paper/shiltsev\_Tev\_vibrations\_v2.pdf}

\bibitem{Aghanim:2018eyx} 
  N.~Aghanim {\it et al.} [Planck Collaboration],
  arXiv:1807.06209 [astro-ph.CO].


\end{thebibliography}
\end{document}